# Mixtures of Spatial Spline Regressions

Hien D. Nguyen, Geoffrey J. McLachlan, and Ian A. Wood

*Department of Mathematics, University of Queensland, St. Lucia, 4072, Australia*

*g.mclachlan@uq.edu.au (Geoffrey J. McLachlan)*

**Abstract**

We present an extension of the functional data analysis framework for univariate functions to the analysis of surfaces: functions of two variables. The spatial spline regression (SSR) approach developed can be used to model surfaces that are sampled over a rectangular domain. Furthermore, combining SSR with linear mixed effects models (LMM) allows for the analysis of populations of surfaces, and combining the joint SSR-LMM method with finite mixture models allows for the analysis of populations of surfaces with sub-family structures.

Through the mixtures of spatial splines regressions (MSSR) approach developed, we present methodologies for clustering surfaces into sub-families, and for performing surface-based discriminant analysis. The effectiveness of our methodologies, as well as the modeling capabilities of the SSR model are assessed through an application to handwritten character recognition.

## 1 Introduction

In recent years, functional data analysis (FDA) (Ramsay and Silverman, 1997), the data analysis of curves, has become a popular and powerful tool for statistical analysis and pattern recognition. With the ability to handle data arising from infinitely dense curves in a greatly reduced dimensionality, FDA has lent itself to numerous modern applications in a large variety of scientific areas. A survey of applications of FDA to biological, economic, medical, machine learning, and sociological data, can be found in Ramsay and Silverman (2002).

Of interest are applications of FDA to the problems of classifying and clustering curvilinear data. Here, we define classification as the supervised learning of discrimination rules based on labeled data from functions belonging to different groups, and clustering as the unsupervised learning of different groups of functions based on unlabeled training data.

Currently, there is a large literature on techniques developed for the clustering and classification of functional data. A short survey of interesting developments in the area include: B-spline regression for classification by linear discriminant analysis (LDA) and clustering by mixtures of linear mixed models (MLMM) (James and Hastie, 2001; James and Sugar, 2003), Fourier basis regression for clustering by MLMM (Ng et al., 2006), nonlinear regression for classifying and clustering by Gaussian mixture models (GMM) (Ma et al., 2002) and MLMM (Wang et al., 2007; Hou et al., 2008), piecewise polynomial regression for classifying and clustering by GMM (Chamroukhi et al., 2010; Chamroukhi and Glotin, 2012), Gaussian process regression for classifying by principal component analysis (Hall et al., 2001) and by centroid-based methods (Delaigle and Hall, 2012), curve classifying by support vector machine (Rossi and Villa, 2006), and curve clustering nonparametric density estimation (Boulle, 2012).

The main focus of the FDA literature has concentrated on the analysis of data generated by univariate functions with a lesser emphasis on exploring the analysis of surface data, generated from bivariate functions. Some important methods for dealing with surface data include: Kriging (Matheron, 1963; Cressie, 1991), thin-plate splines (Duchon, 1977; Wahba, 1990), multivariate adaptive regression splines (Friedman, 1991), soap film smoothing (Wood et al., 2008), and spatial spline regression (SSR) models (Malfait and Ramsay, 2003; Ramsay et al., 2011; Sangalli et al., 2013). The developments in bivariate FDA have concentrated on smoothing and accurate estimation based on data generated from a single function rather than inference regarding data from multiple functions.

In this article, we extend the work on functional clustering by James and Sugar (2003) and functional discriminant analysis by James and Hastie (2001) to the case of bivariate functional data. We do this by presenting a novel application of bivariate spline functions, as used in SSR, to the problem of estimating the probability distribution of surfaces through multiple sets of functional observations. This is approached by using linear mixed effects models (LMM) and naturally leads to clustering by MLMM (Celeux et al., 2005; James and Sugar, 2003; Ng et al., 2006), where we assume that the observed samples arise from a mixture of generative functions. We call this approach mixtures of spatial spline regressions (MSSR). MSSR models not only allow for the clustering of surfaces into groups, but also lets us estimate mixture distributions for different classes of observations in supervised data. When observations are obtained along with their class labels, MSSR can be used to estimate the posterior probability of membership of the different surface classes observed. These probability distributions can then be used for classification by the approach we term mixtures of spatial spline regressions discriminant analysis (MSSRDA). We note that MSSRDA is a bivariate, spatial analog of the mixture discriminant analysis (Hastie and Tibshirani, 1996) extension of the LDA approach (James and Hastie, 2001) for univariate functional data.

This paper is organized as follows. We present the SSR model for a single surface and the LMM extension for describing the distribution of bivariate functional data sets arising from perturbations of a single generative surface in **Section 2**. In **Section 3**, we describe the MSSR model for clustering when the the functional data arise from a mixture distribution of generative surfaces and, in **Section 4**, we extend the use of MSSR to classification through MSSRDA. In **Section 5**, we demonstrate the application of MSSR and MSSRDA to the ZIP code data set from Hastie et al. (2009), a subset of the MNIST data set (Le Cun et al., 1990), and discuss our results. Finally, in **Section 6**, we provide conclusions, remarks, and an assessment of possible future directions for this research.

## 2 Spatial spline regressions

In this section, we discuss the use of SSR for the problem of fitting a model to data observed from a single surface within a rectangular domain. The technique is then extended to the fitting of data generated from a family of generative surfaces by using LMM.

Let $\boldsymbol{w}_k = \left(y_k, \boldsymbol{x}_k^T\right)^T$ $(k = 1, .., m)$ be a sample of $m$ observations. Here $\boldsymbol{x}_k = (x_{1k}, x_{2k})^T$ and $y_k = \mu(\boldsymbol{x}_k) + e_k$, where $\mu$ is an unknown bivariate function and the superscript $T$ denotes transposition. We make the closed rectangular domain restriction that $(x_{1k}, x_{2k}) \in R$, where $R = \left[x_1^-, x_1^+\right] \times \left[x_2^-, x_2^+\right]$ with $x_1^- < x_1^+$ and $x_2^- < x_2^+$ , and let $y_k$ be an observation of the function $\mu$ evaluated at the point $\boldsymbol{x}_k$ with an additive white noise component $e_k \sim N(0, \sigma^2)$.

If $\mu$ has a known parametric form, it is possible to estimate $\mu$ to get an optimal least-squares or maximum-likelihood fitted function $\hat{\mu}(\boldsymbol{x})$. This can be achieved with nonlinear regression analysis (Bates and Watts, 1988). Nonlinear regression analysis is not possible for unknown $\mu$.

### *2.1 Nodal basis functions*

In the univariate FDA literature, a popular approximation of $\mu$ for observations that are in a bounded set is to use B-splines (Ramsay and Silverman, 1997). The idea of univariate B-splines can be extended to bivariate surface estimation through the use of nodal basis functions (NBFs) from the finite element method (FEM) literature (Braess, 2001). The idea of using NBFs was first introduced by Malfait and Ramsay (2003) for estimation of historical functional data, and later used in Ramsay et al. (2011) and Sangalli et al. (2013) for estimation of sparsely sampled surfaces with applications in demography and hemodynamics; the study of blood flow and circulation.

Like univariate B-spline interpolation, a decision must first be made on the number of basis functions to use to estimate the function $\mu$. We take a regular grid approach, reminiscent of Malfait and Ramsay (2003), and define $d = d_1 d_2$ as the number of bases used where $d_1$ and $d_2$ are the number of columns and rows of bases, respectively. The domain $R$ is then divided into $d$ regularly spaced points $c_l$ for $l = 1, ..., d$ such that

$$\begin{aligned}
\boldsymbol{c}_1 &= \left(x_1^-, x_2^-\right), \\
\boldsymbol{c}_2 &= \left(x_1^- + \delta_1, x_2^-\right), \\
&\vdots \\
\boldsymbol{c}_{d_1-1} &= \left(x_1^+ - \delta_1, x_2^-\right), \\
\boldsymbol{c}_{d_1} &= \left(x_1^+, x_2^-\right), \\
\boldsymbol{c}_{d_1+1} &= \left(x_1^-, x_2^- + \delta_2\right), \\
&\vdots \\
\boldsymbol{c}_{2d_1-1} &= \left(x_1^+ - \delta_1, x_2^- + \delta_2\right), \\
\boldsymbol{c}_{2d_1} &= \left(x_1^+, x_2^- + \delta_2\right), \\
\boldsymbol{c}_{2d_1+1} &= \left(x_1^-, x_2^- + 2\delta_2\right), \\
&\vdots \\
\boldsymbol{c}_{d-1} &= \left(x_1^+ - \delta_1, x_2^+\right), \\
\boldsymbol{c}_{d} &= \left(x_1^+, x_2^+\right),
\end{aligned} \tag{1}$$

where $\delta_1 = \left(x_1^+ - x_1^-\right) / (d_1 - 1)$ and $\delta_2 = \left(x_2^+ - x_2^-\right) / (d_2 - 1)$.

At every point $\boldsymbol{c}_l \in C$, as given by (1), we place a piecewise linear Lagrange triangular finite element NBF (Sangalli et al., 2013) of the form

$$
s\left(\boldsymbol{x};\boldsymbol{c},\delta_1,\delta_2\right)=\begin{cases}
-\frac{x_2}{\delta_2}+\frac{x_{2c}\delta_2}{\delta_2} & \text{if } \boldsymbol{x}\in\left\{(x_1,x_2): x_{1c}<x_1\leq x_{1c}+\delta_1,\, \frac{\delta_2}{\delta_1}x_1+\frac{\delta_1 x_{2c}-\delta_2 x_{1c}}{\delta_1}\leq x_2\leq x_{2c}+\delta_2\right\}, \\
-\frac{x_1}{\delta_1}+\frac{x_{1c}+\delta_1}{\delta_1} & \text{if } \boldsymbol{x}\in\left\{(x_1,x_2): x_{1c}<x_1\leq x_{1c}+\delta_1, x_{2c}\leq x_2<\frac{\delta_2}{\delta_1}x+\frac{\delta_1 x_{2c}-\delta_2 x_{1c}}{\delta_1}\right\}, \\
-\frac{x_1}{\delta_1}+\frac{x_2}{\delta_2}+\frac{\delta_1\delta_2+\delta_2 x_{1c}-\delta_1 x_{2c}}{\delta_1\delta_2} & \text{if } \boldsymbol{x}\in\left\{(x_1,x_2): x_{1c}\leq x_1\leq x_{1c}+\delta_1,\, \frac{\delta_2}{\delta_1}x_1+\frac{\delta_1 x_{2c}-\delta_2 x_{1c}-\delta_1\delta_2}{\delta_1}\leq x_2<x_{2c}\right\}, \\
\frac{x_2}{\delta_2}+\frac{\delta_2-x_{2c}}{\delta_2} & \text{if } \boldsymbol{x}\in\left\{(x_1,x_2): x_{1c}-\delta_1\leq x_1<x_{1c}, x_{2c}-\delta_2\leq, x_2\leq\frac{\delta_2}{\delta_1}x_1+\frac{\delta_1 x_{2c}-\delta_2 x_{1c}}{\delta_1}\right\}, \\
\frac{x_1}{\delta_1}+\frac{\delta_1-x_{1c}}{\delta_1} & \text{if } \boldsymbol{x}\in\left\{(x_1,x_2): x_{1c}-\delta_1\leq x_1<x_{1c},\, \frac{\delta_2}{\delta_1}x_1+\frac{\delta_1 x_{2c}-\delta_2 x_{1c}}{\delta_1}<x_2\leq x_{2c}\right\}, \\
\frac{x_1}{\delta_1}-\frac{x_2}{\delta_2}+\frac{\delta_1\delta_2+\delta_1 x_{2c}-\delta_2 x_{1c}}{\delta_1\delta_2} & \text{if } \boldsymbol{x}\in\left\{(x_1,x_2): x_{1c}-\delta_1\leq x_1\leq x_{1c}, x_{2c}<x_2\leq\frac{\delta_2}{\delta_1}x_1+\frac{\delta_1 x_{2c}+\delta_1\delta_2-\delta_2 x_{1c}}{\delta_1}\right\}, \\
0 & \text{otherwise,}
\end{cases} \tag{2}
$$

where $\boldsymbol{x}\in R$ and $\boldsymbol{c}=(x_{1c},x_{2c})\in\{\boldsymbol{c}_1,...,\boldsymbol{c}_d\}$. The NBF will be shortened to $s\left(\boldsymbol{x};\boldsymbol{c}\right)$ for ease of notation due to the constancy of $\delta_1$ and $\delta_2$. We give an example of an NBF in **Figure 1**.

Using (2), we can approximate the function $\mu$ within the domain $\boldsymbol{x}\in R$ by a linear combination of the NBFs,

$$
\mu\left(\boldsymbol{x}\right)=\sum_{l=1}^{d}\beta_l s\left(\boldsymbol{x};\boldsymbol{c}_l\right), \tag{3}
$$

where $\boldsymbol{\beta}=\left(\beta_1,...,\beta_d\right)^T$ is a vector of basis coefficients. It can also be shown that the approximation scheme (3) can exactly interpolate $\mu$ if $\mu$ belongs to the family of affine planes (Malfait and Ramsay, 2003).

### *2.2 Fitting SSR models*

The estimation problem of fitting a surface to the sample $\boldsymbol{w}_c=\left(\boldsymbol{w}_1^T,...,\boldsymbol{w}_m^T\right)^T$ can be approached by approximating $\mu$ by (3), and assuming that

$$
y_k=\sum_{l=1}^{d}\beta_l s\left(\boldsymbol{x}_k;\boldsymbol{c}_l\right)+e_k \tag{4}
$$

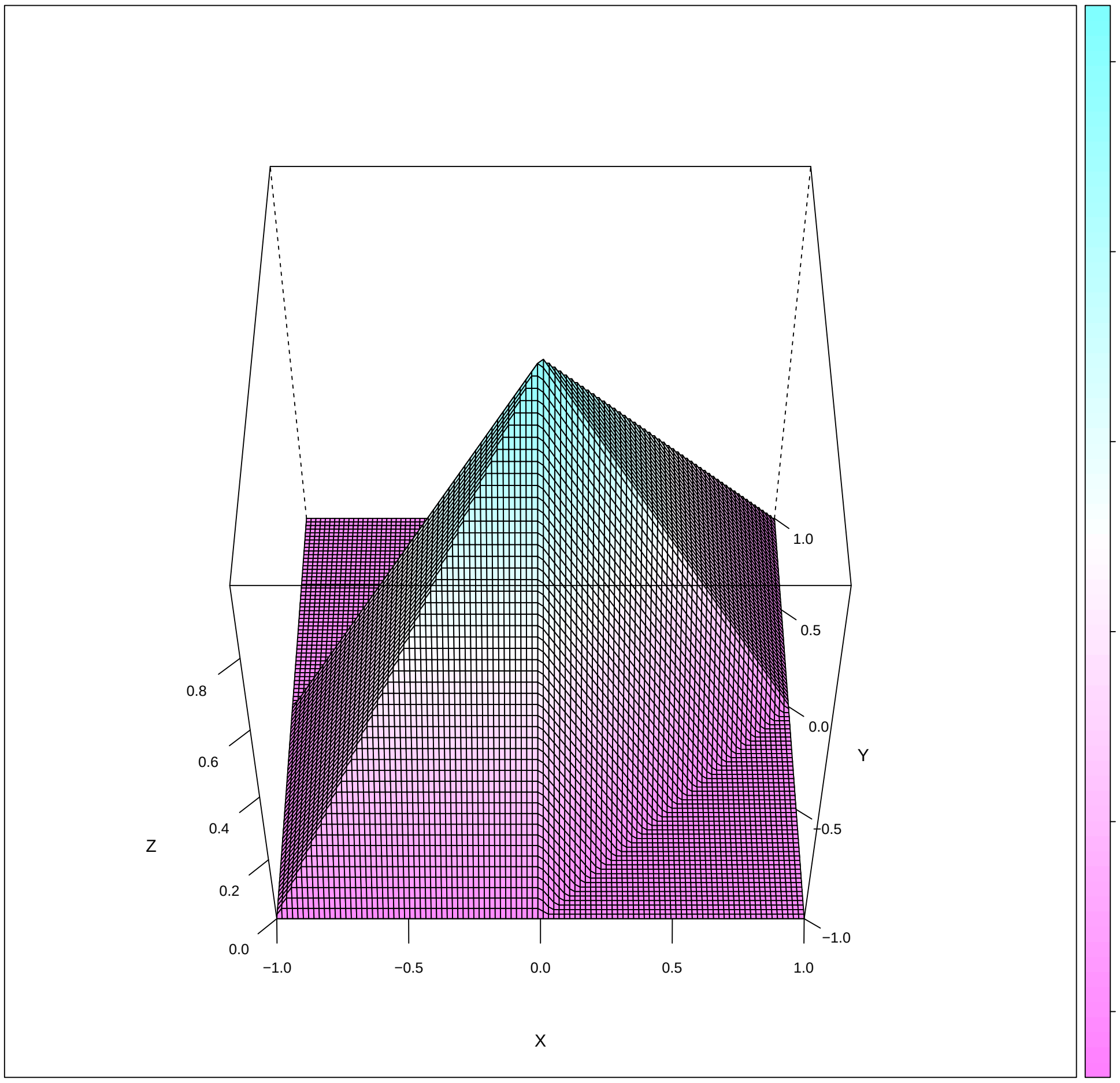

Figure 1. Nodal basis function $s(\boldsymbol{x}; \boldsymbol{c}, \delta_1, \delta_2)$, where $\boldsymbol{c} = (0, 0)$, $\delta_1 = 1$ and $\delta_2 = 1$.

for all $k = 1, ..., m$. This leads to the matrix form

$$\boldsymbol{y} = \boldsymbol{S}\boldsymbol{\beta} + \boldsymbol{e}, \tag{5}$$

where $\boldsymbol{y} = (y_1, ..., y_m)^T$, $\boldsymbol{e} = (e_1, ..., e_m)^T$ is a Gaussian error distributed as $N(\mathbf{0}_m, \sigma^2 \boldsymbol{I}_m)$, and where

$$
\boldsymbol{S} = \begin{bmatrix} s\left(\boldsymbol{x}_1; \boldsymbol{c}_1\right) & s\left(\boldsymbol{x}_1; \boldsymbol{c}_2\right) & \cdots & s\left(\boldsymbol{x}_1; \boldsymbol{c}_d\right) \\ s\left(\boldsymbol{x}_2; \boldsymbol{c}_1\right) & s\left(\boldsymbol{x}_2; \boldsymbol{c}_2\right) & \cdots & s\left(\boldsymbol{x}_2; \boldsymbol{c}_d\right) \\ \vdots & \vdots & \ddots & \vdots \\ s\left(\boldsymbol{x}_m; \boldsymbol{c}_1\right) & s\left(\boldsymbol{x}_m; \boldsymbol{c}_2\right) & \cdots & s\left(\boldsymbol{x}_m; \boldsymbol{c}_d\right) \end{bmatrix} \tag{6}
$$

and $\boldsymbol{I}_m$ is the identity matrix of subscripted dimension.

The probability density function of $\boldsymbol{y}$ is given by

$$
f\left(\boldsymbol{y}; \boldsymbol{\beta}, \sigma^2\right) = \phi\left(\boldsymbol{y}; \boldsymbol{S\beta}, \sigma^2 \boldsymbol{I}_m\right), \tag{7}
$$

where $\phi$ is the $m$-variate Gaussian probability density function. The likelihood function for $\boldsymbol{\beta}$ and $\sigma^2$ is given by

$$
L\left(\boldsymbol{\beta}, \sigma^2\right) = f\left(\boldsymbol{y}; \boldsymbol{\beta}, \sigma^2\right) \tag{8}
$$

under the linear model (5).

Using the likelihood function (8), the maximum likelihood estimates (MLE) for the model parameters are given by

$$
\hat{\boldsymbol{\beta}} = \left(\boldsymbol{S}^T \boldsymbol{S}\right)^{-1} \boldsymbol{S}^T \boldsymbol{y} \tag{9}
$$

and

$$
\hat{\sigma}^2 = \frac{\left(\boldsymbol{y} - \boldsymbol{S\beta}\right)^T \left(\boldsymbol{y} - \boldsymbol{S\beta}\right)}{m}, \tag{10}
$$

where $\hat{\boldsymbol{\beta}}$ and $\hat{\sigma}^2$ is the MLE for $\boldsymbol{\beta}$ and $\sigma^2$, respectively. The estimate $\hat{\boldsymbol{\beta}}$ can be substituted back into (3) in order to interpolate the function $\mu$ and is noted to be equivalent to a least-squares fitting of SSR models (Sangalli et al., 2013).

The SSR presented so far closely resembles the results from Sangalli et al. (2013) and can be extended upon to make inferences about the fitted surface. However, the SSR framework does not allow for the analysis of data generated from a family of surfaces where each surface is generated from the same distribution. We therefore introduce the spatial spline regression linear mixed effects model (SSR-LMM) in order to handle such a scenario.

### *2.3 Linear mixed effects models*

Instead of a single observed sample from a single surface, we now consider the case of $n$ samples of observations from distinct surfaces $\boldsymbol{w}_{cj} = \left(\boldsymbol{w}_{j1}^T, ..., \boldsymbol{w}_{jm_j}^T\right)^T$ for $j = 1, ..., n$. For each $j$, $\boldsymbol{w}_{jk} = \left(y_{jk}, \boldsymbol{x}_{jk}^T\right)^T$, where $m_j$ is the number of points observed for surface $j$, and $y_{jk} = \mu_j\left(\boldsymbol{x}_{jk}\right) + e_{jk}$, where $e_{jk} \sim N\left(0, \sigma^2\right)$ for all $j = 1, ..., n$ and $k = 1, ...m_j$. The function $\mu_j$ is an unknown function such that $E\left(\mu_j\left(\boldsymbol{x}\right)\right) = \mu\left(\boldsymbol{x}\right)$ for all $j$ for some mean function $\mu$. If the distribution of $\mu_j$ were known, it would be possible to estimate the function $\mu$ using nonlinear mixed effects models (Pinheiro and Bates, 2000). The unknown nature of $\mu$ once again necessitates approximation, for which we use SSR.

For each sample $\boldsymbol{w}_{cj}$, we can postulate corresponding to (3) that

$$\mu_j\left(\boldsymbol{x}\right) = \sum_{l=1}^{d}\left(\beta_l + b_{jl}\right) s\left(\boldsymbol{x}; \boldsymbol{c}_l\right), \tag{11}$$

where $b_{jl} \sim N\left(0, \xi^2\right)$ are random effects. We then have that

$$\boldsymbol{y}_j = \boldsymbol{S}_j \left(\boldsymbol{\beta} + \boldsymbol{b}_j\right) + \boldsymbol{e}_j, \tag{12}$$

where

$$\boldsymbol{S}_j = \begin{bmatrix} s\left(\boldsymbol{x}_{j1}; \boldsymbol{c}_1\right) & s\left(\boldsymbol{x}_{j1}; \boldsymbol{c}_2\right) & \cdots & s\left(\boldsymbol{x}_{j1}; \boldsymbol{c}_d\right) \\ s\left(\boldsymbol{x}_{j2}; \boldsymbol{c}_1\right) & s\left(\boldsymbol{x}_{j2}; \boldsymbol{c}_2\right) & \cdots & s\left(\boldsymbol{x}_{j2}; \boldsymbol{c}_d\right) \\ \vdots & \vdots & \ddots & \vdots \\ s\left(\boldsymbol{x}_{jm_j}; \boldsymbol{c}_1\right) & s\left(\boldsymbol{x}_{jm_j}; \boldsymbol{c}_2\right) & \cdots & s\left(\boldsymbol{x}_{jm_j}; \boldsymbol{c}_d\right) \end{bmatrix}, \tag{13}$$

$\boldsymbol{y}_j = \left(y_{j1}, ..., y_{jm_j}\right)^T$, $\boldsymbol{e}_j = \left(e_{j1}, ..., e_{jm_j}\right)^T$ and $\boldsymbol{b}_j = \left(b_{j1}, ..., b_{jd}\right)^T$. Here $\boldsymbol{e}_j \sim N\left(\mathbf{0}_{m_j}, \sigma^2 \boldsymbol{I}_{m_j}\right)$ and $\boldsymbol{b}_j \sim N\left(\mathbf{0}_d, \xi^2 \boldsymbol{I}_d\right)$, where $\mathbf{0}_d$ denotes the zero vector of the subscripted dimension.

The density function for each $\boldsymbol{y}_j$ can be expressed as

$$f\left(\boldsymbol{y}_j; \boldsymbol{\Psi}, \boldsymbol{x}_j\right) = \phi\left(\boldsymbol{y}_j; \boldsymbol{S}_j\boldsymbol{\beta}, \xi^2 \boldsymbol{S}_j \boldsymbol{S}_j^T + \sigma^2 \boldsymbol{I}_{m_j}\right) \tag{14}$$

and thus we can write the likelihood function for the parameter vector $\boldsymbol{\Psi} = \left(\boldsymbol{\beta}^T, \sigma^2, \xi^2\right)^T$ as

$$L\left(\boldsymbol{\Psi}\right) = \prod_{j=1}^{n} \phi\left(\boldsymbol{y}_j; \boldsymbol{S}_j\boldsymbol{\beta}, \xi^2 \boldsymbol{S}_j \boldsymbol{S}_j^T + \sigma^2 \boldsymbol{I}_{m_j}\right). \tag{15}$$

### *2.4 Fitting SSR-LMM*

The MLE of the parameter vector $\boldsymbol{\Psi}$ can be found by maximizing (15) over $\boldsymbol{\Psi}$. The maximization can be achieved with numerical methods such as by using Newton or Quasi-Newton type algorithms (McLachlan and Krishnan, 2008), which can become numerically intensive given either a large number of samples $n$, number of observations $m_j$ or dimensionality of the spline approximation $d$. We can instead consider that the random effects $\boldsymbol{b}_j$ are latent variables which can be interpreted as missing data. In this case, we can use the EM algorithm (Dempster et al., 1977; McLachlan and Krishnan, 2008). In the EM framework, the complete-data likelihood is given by

$$L_c(\boldsymbol{\Psi}) = \prod_{j=1}^{n} \phi\left(\boldsymbol{y}_j; \boldsymbol{S}_j\boldsymbol{\beta}, \sigma^2\boldsymbol{I}_{m_j}\right) \phi\left(\boldsymbol{b}_j; \boldsymbol{0}, \xi^2\boldsymbol{I}_d\right). \tag{16}$$

Using (16) we can devise an EM algorithm for obtaining the MLE of $\boldsymbol{\Psi}$ as well as estimate the missing data vectors $\boldsymbol{b}_j$. Starting with an initial estimate $\boldsymbol{\Psi}^{(0)} = \left(\boldsymbol{\beta}^{(0)T}, \sigma^{(0)^2}, \xi^{(0)^2}\right)^T$ and letting $\boldsymbol{\Psi}^{(k)} = \left(\boldsymbol{\beta}^{(k)T}, \sigma^{(k)^2}, \xi^{(k)^2}\right)^T$, the E- and M-steps of the algorithm can be described as follows.

On the $(k+1)$ th E-step, the required conditional expectations $E_{\boldsymbol{\Psi}^{(k)}}(\boldsymbol{b}_j|\boldsymbol{w}_{cj})$, $E_{\boldsymbol{\Psi}^{(k)}}\left(\boldsymbol{b}_j^T\boldsymbol{b}_j|\boldsymbol{w}_{cj}\right)$, $E_{\boldsymbol{\Psi}^{(k)}}(\boldsymbol{e}_j|\boldsymbol{w}_{cj})$, and $E_{\boldsymbol{\Psi}^{(k)}}\left(\boldsymbol{e}_j^T\boldsymbol{e}_j|\boldsymbol{w}_{cj}\right)$ are computed as

$$E_{\boldsymbol{\Psi}^{(k)}}(\boldsymbol{b}_j|\boldsymbol{w}_{cj}) = \xi^{(k)^2}\boldsymbol{S}_j^T\left(\xi^{(k)^2}\boldsymbol{S}_j\boldsymbol{S}_j^T + \sigma^{(k)^2}\boldsymbol{I}_{m_j}\right)^{-1}\left(\boldsymbol{y}_j - \boldsymbol{S}_j\boldsymbol{\beta}^{(k)}\right), \tag{17}$$

$$\begin{aligned} E_{\boldsymbol{\Psi}^{(k)}}\left(\boldsymbol{b}_j^T\boldsymbol{b}_j|\boldsymbol{w}_{cj}\right) = &\operatorname{tr}\left(\delta^{2(k)}\left(\boldsymbol{I}_d - \xi^{2(k)}\boldsymbol{S}_j^T\left(\xi^{2(k)}\boldsymbol{S}_j\boldsymbol{S}_j^T + \sigma^{2(k)}\boldsymbol{I}_{m_j}\right)^{-1}\boldsymbol{S}_j\right)\right) \\ &+E_{\boldsymbol{\Psi}^{(k)}}(\boldsymbol{b}_j|\boldsymbol{w}_{cj})^T E_{\boldsymbol{\Psi}^{(k)}}(\boldsymbol{b}_j|\boldsymbol{w}_{cj}), \end{aligned} \tag{18}$$

$$E_{\boldsymbol{\Psi}^{(k)}}(\boldsymbol{e}_j|\boldsymbol{w}_{cj}) = \boldsymbol{y}_j - \boldsymbol{S}_j\left(\boldsymbol{\beta} + E_{\boldsymbol{\Psi}^{(k)}}(\boldsymbol{b}_j|\boldsymbol{w}_{cj})\right), \tag{19}$$

and

$$E_{\mathbf{\Psi}^{(k)}}\left(\boldsymbol{e}_j^T\boldsymbol{e}_j|\boldsymbol{w}_{cj}\right) = \mathrm{tr}\left(\xi^{2(k)}\boldsymbol{S}_j\left(\boldsymbol{I}_d - \xi^{2(k)}\boldsymbol{S}_j^T\left(\xi^{2(k)}\boldsymbol{S}_j\boldsymbol{S}_j^T + \sigma^{2(k)}\boldsymbol{I}_{m_j}\right)^{-1}\boldsymbol{S}_j\right)\boldsymbol{S}_j^T\right) + E_{\mathbf{\Psi}^{(k)}}\left(\boldsymbol{e}_j|\boldsymbol{w}_{cj}\right)^T E_{\mathbf{\Psi}^{(k)}}\left(\boldsymbol{e}_j|\boldsymbol{w}_{cj}\right) \tag{20}$$

for each $j = 1, ..., n$.

If we let

$$E_{\mathbf{\Psi}^{(k)}}\left(\boldsymbol{b}_j|\boldsymbol{w}_{cj}\right) = \boldsymbol{b}_j^{(k+1)}, \tag{21}$$

$$\lambda_{\boldsymbol{b}_j}^{(k)} = \mathrm{tr}\left(\xi^{(k)^2}\left(\boldsymbol{I}_d - \xi^{(k)^2}\boldsymbol{S}_j^T\left(\xi^{(k)^2}\boldsymbol{S}_j\boldsymbol{S}_j^T + \sigma^{(k)^2}\boldsymbol{I}_{m_j}\right)^{-1}\boldsymbol{S}_j\right)\right), \tag{22}$$

and

$$\lambda_{\boldsymbol{e}_j}^{(k)} = \mathrm{tr}\left(\xi^{(k)^2}\boldsymbol{S}_j\left(\boldsymbol{I}_d - \xi^{(k)^2}\boldsymbol{S}_j^T\left(\xi^{(k)^2}\boldsymbol{S}_j\boldsymbol{S}_j^T + \sigma^{(k)^2}\boldsymbol{I}_{m_j}\right)^{-1}\boldsymbol{S}_j\right)\boldsymbol{S}_j^T\right), \tag{23}$$

we can compute the $(k+1)$ th iteration of the M-step estimate $\mathbf{\Psi}^{(k+1)}$ as

$$\boldsymbol{\beta}^{(k+1)} = \left(\sum_{j=1}^{n}\boldsymbol{S}_j^T\boldsymbol{S}_j\right)^{-1}\left(\sum_{j=1}^{n}\boldsymbol{S}_j^T\left(\boldsymbol{y}_j - \boldsymbol{S}_j\boldsymbol{b}_j^{(k+1)}\right)\right), \tag{24}$$

$$\sigma^{(k+1)^2} = \frac{\sum_{j=1}^{n}\left(\left(\boldsymbol{y}_j - \boldsymbol{S}_j\left(\boldsymbol{\beta}+\boldsymbol{b}_j^{(k+1)}\right)\right)^T\left(\boldsymbol{y}_j - \boldsymbol{S}_j\left(\boldsymbol{\beta}+\boldsymbol{b}_j^{(k+1)}\right)\right) + \lambda_{\boldsymbol{e}_j}\right)}{\sum_{j=1}^{n} m_j}, \tag{25}$$

and

$$\xi^{(k+1)^2} = \frac{\sum_{j=1}^{n}\left(\boldsymbol{b}_j^{(k+1)T}\boldsymbol{b}_j^{(k+1)} + \lambda_{\boldsymbol{b}_j}\right)}{nd}. \tag{26}$$

The E- and the M-steps are alternated repeatedly until the increase in likelihood evaluations falls below a numerical threshold. The final iterate of the parameter vector $\hat{\boldsymbol{\Psi}} = \left(\hat{\boldsymbol{\beta}}, \hat{\sigma}^2, \hat{\xi}^2\right)^T$ can then be used as the MLE of the system of equations (12), and the conditional expectation $\hat{\boldsymbol{b}}_j = E_{\hat{\boldsymbol{\Psi}}^{(k)}}\left(\boldsymbol{b}_j|\boldsymbol{w}_{cj}\right)$ can be used as an estimate of $\boldsymbol{b}_j$. Using the MLE $\hat{\boldsymbol{\Psi}}$, $\mu$ can be approximated as $\sum_{l=1}^{d}\hat{\beta}_l s\left(\boldsymbol{x};\boldsymbol{c}_l\right)$ and $\mu_j$ can be approximated as $\sum_{l=1}^{d}\left(\hat{\beta}_l + \hat{b}_{jl}\right) s\left(\boldsymbol{x};\boldsymbol{c}_l\right)$ for each surface $j$. Note that our estimates $\hat{\sigma}^2$ and $\hat{\xi}^2$ also allow us to describe the nature of the additive white noise $\boldsymbol{e}_j$ for each surface, as well as the nature of the distribution of the $n$ surfaces around the average $\sum_{l=1}^{d}\hat{\beta}_l s\left(\boldsymbol{x};\boldsymbol{c}_l\right)$.

## 3 Mixtures of spatial spline regressions

So far, we have described the SSR-LMM approach that allows us to estimate and make inferences based on data sets for surfaces from a common family. In this section, we extend the SSR-LMM to model the situation where the $n$ observations $\boldsymbol{w}_{cj}$ $(j = 1, ..., n)$ are generated from $g$ different families of surfaces through the use of a $g$ component mixture model (McLachlan and Basford, 1988; McLachlan and Peel, 2000). We suppose that the $j$th observation is generated from an unknown function $\mu_{ij}$, with the functions $\mu_{ij}$ $(j = 1, ..., n)$ distributed around some mean function $\mu_i$, where $E\left(\mu_{ij}\left(\boldsymbol{x}\right)\right) = \mu_i$ for each component $i$ $(i = 1, ..., g)$. If the observed sample $\boldsymbol{w}_{cj}$ has been drawn from the $i$th component of the mixture mode, we can write $y_{jk} = \mu_{ij}\left(\boldsymbol{x}_{jk}\right) + e_{ijk}$, where $e_{ijk} \sim N\left(0, \sigma^2\right)$ is an additive white noise component.

### 3.1 Fitting MSSR Models

Like the models SSR and SSR-LMM, if the functions $\mu_i$ are known, nonlinear MLMM can be used to estimate the parameters directly (Wang et al., 2007; Hou et al., 2008). Given that we do not know the form of $\mu_i$, once again, we can use SSR to make an approximation. Supposing that the surface for each observation $j$ $(j = 1, ..., n)$, conditioned on $\boldsymbol{w}_{cj}$ belonging to component $i$ $(i = 1, ..., g)$, can be approximated by

$$\mu_{ij}(\boldsymbol{x}) = \sum_{l=1}^{d} (\beta_{il} + b_{ijl})\, s(\boldsymbol{x}; \boldsymbol{c}_l), \tag{27}$$

where $\boldsymbol{\beta}_i = (\beta_{i1}, ..., \beta_{id})^T$ and $\boldsymbol{b}_{ij} = (b_{ij1}, ..., b_{ijd})^T \sim N(\boldsymbol{0}_d, \xi_i^2 \boldsymbol{I}_d)$ are the basis coefficients and random effects, respectively.

The approximation (27) gives us the matrix form

$$\boldsymbol{y}_j = \boldsymbol{S}_j(\boldsymbol{\beta}_i + \boldsymbol{b}_{ij}) + \boldsymbol{e}_{ij}, \tag{28}$$

where $\boldsymbol{e}_{ij} \sim N\left(\boldsymbol{0}_{m_j}, \sigma^2 \boldsymbol{I}_{m_j}\right)$ for each observation $\boldsymbol{w}_{cj}$ $(j = 1, ..., n)$, given membership of component $i$ $(i = 1, ..., g)$.

If we let the probability of membership in the $i$th component be $\pi_i$ such that $\sum_{i=1}^{g} \pi_i = 1$, we can write the probability density function of $\boldsymbol{y}_j$ as

$$f(\boldsymbol{y}_j; \boldsymbol{\Psi}, \boldsymbol{x}_j) = \sum_{i=1}^{g} \pi_i \phi\left(\boldsymbol{y}_j; \boldsymbol{S}_j \boldsymbol{\beta}_i, \xi_i^2 \boldsymbol{S}_j \boldsymbol{S}_j^T + \sigma^2 \boldsymbol{I}_{m_j}\right), \tag{29}$$

where $\boldsymbol{\Psi} = \left(\sigma^2, \boldsymbol{\beta}_1^T, ..., \boldsymbol{\beta}_g^T, \xi_1^2, ...\xi_g^2, \boldsymbol{\pi}^T\right)^T$ and $\boldsymbol{\pi} = (\pi_1, ..., \pi_g)^T$.

We observe that the density function $f\left(\boldsymbol{y}_j; \boldsymbol{\Psi}\right)$ is a Gaussian mixture model with $g$ components which we can use to write the likelihood function across all $n$ observations $\boldsymbol{w}_{cj}$ $(j = 1, ..., n)$ as

$$L\left(\boldsymbol{\Psi}\right) = \prod_{j=1}^{n} \sum_{i=1}^{g} \pi_i \phi\left(\boldsymbol{y}_j; \boldsymbol{S}_j \boldsymbol{\beta}_i, \xi_i^2 \boldsymbol{S}_j \boldsymbol{S}_j^T + \sigma^2 \boldsymbol{I}_{m_j}\right). \tag{30}$$

Similarly to (15), maximizing (30) over the parameter $\boldsymbol{\Psi}$ is a difficult numerical problem. As an alternative, we can write the complete-data likelihood function by introducing latent variables $\boldsymbol{z}_j$ for $j = 1, ..., n$, where $\boldsymbol{z}_j = \left(Z_{1j}, ..., Z_{gj}\right)^T \sim \text{Multinomial}\left(1, \boldsymbol{\pi}\right)$ and

$$Z_{ij} = \begin{cases} 1, & \text{if observation } \boldsymbol{w}_{cj} \text{belongs to component } i, \\ 0, & \text{otherwise.} \end{cases} \tag{31}$$

Using (31) and incorporating the random effects $\boldsymbol{b}_{ij}$ into the model as missing data in our EM framework, we can express the complete-data likelihood function based on the observations $\boldsymbol{w}_{cj}$ $(j = 1, ..., n)$ as

$$L_c\left(\boldsymbol{\Psi}\right) = \prod_{j=1}^{n} \prod_{i=1}^{g} \left(\pi_i \phi\left(\boldsymbol{y}_j; \boldsymbol{S}_j\left(\boldsymbol{\beta}_i + \boldsymbol{b}_{ij}\right), \sigma^2 \boldsymbol{I}_{m_j}\right) \phi\left(\boldsymbol{b}_{ij}; \boldsymbol{0}_d, \xi_i^2 \boldsymbol{I}_d\right)\right)^{z_{ij}}. \tag{32}$$

We can work with the complete-data likelihood $L_c\left(\boldsymbol{\Psi}\right)$, via the EM algorithm, to maximize the incomplete-data likelihood $L\left(\boldsymbol{\Psi}\right)$. See **Appendix**.

The E- and M-steps are alternated repeatedly until the increase in consecutive likelihood evaluations falls below a numerical threshold to give $\hat{\boldsymbol{\Psi}} = \left(\hat{\sigma}^2, \hat{\boldsymbol{\beta}}_1^T, ..., \hat{\boldsymbol{\beta}}_g^T, \hat{\xi}_1^2, ..., \hat{\xi}_g^2, \hat{\boldsymbol{\pi}}^T\right)^T$ which can be used to make inferences regarding the different families of surfaces. As with the SSR-LMM model, we can also obtain a description of the distribution of the surfaces around the family average surfaces as well as that of the white noise through the estimates $\hat{\xi}_i^2$ for each $i = 1, ..., g$, and $\hat{\sigma}^2$, respectively. We note that SSR-LMM is a special case of MSSR when $g = 1$ and that our method for estimation of the MSSR model resembles the approaches of Celeux et al. (2005)

and Ng et al. (2006).

### *3.2 Clustering with MSSR*

For the EM process for maximizing equation (32), we also need the conditional expectations $\hat{\tau}_{ij} = E_{\hat{\boldsymbol{\Psi}}}\left(Z_{ij}|\boldsymbol{w}_{cj}\right)$ for each $j = 1, ..., n$ and $i = 1, ..., g$. These conditional expectations are in fact conditional probabilities of each observation $\boldsymbol{w}_{cj}$ belonging to component $i$: $\text{pr}_{\hat{\boldsymbol{\Psi}}}\left(Z_{ij} = 1|\boldsymbol{w}_{cj}\right)$. If we consider that each cluster corresponds to a component of the mixture model as defined by (31), then using a decision-theoretic approach, we can allocate each of the observations $\boldsymbol{w}_{cj}$ to one of the $g$ clusters using the plug-in sample version of Bayes' rule (McLachlan, 1992). That is, $\hat{z}_{ij} = 1$, if

$$i = \underset{i'=1,...,g}{\arg\max}\ \hat{\tau}_{i'j}, \tag{33}$$

where $\hat{z}_{ij}$ is an estimate of $Z_{ij}$.

### *3.3 Selecting the number of components g*

In the discussion so far, it has been assumed that the number of families $g$ is given. However, in practice, there are few situations where the number of families is known beforehand.

Since the number $g$ cannot be estimated within the MSSR estimation process, an external method must be devised in order to determine $g$. We suggest that information-theoretic methods can be applied in this situation such as the Bayesian information criterion (BIC) (Schwarz, 1978).

If we let $\gamma$ be a whole number, and $\hat{\boldsymbol{\Psi}}_\gamma = \left(\hat{\sigma}^2, \hat{\boldsymbol{\beta}}_1^T, ..., \hat{\boldsymbol{\beta}}_\gamma^T, \hat{\xi}_1^2, ...\hat{\xi}_\gamma^2, \hat{\boldsymbol{\pi}}^T\right)^T$ be the MLE of the parameter vector with $\gamma$ components, then with the BIC, we can choose

$$g = \arg\min_{\gamma} \ -2\,\log\,L_c\left(\hat{\boldsymbol{\Psi}}_\gamma\right) + \gamma\,(d+2)\log\,n. \tag{34}$$

Since a search over all possibilities is generally not a feasible strategy, other search methods are required to find the optimal value for $g$. Obvious candidates for the purpose are simple, greedy algorithms such as forward selection methods whereby $\gamma$ is either increased until the criterion ceases to improve. The forward selection strategy is implemented in **Section 5**.

## 4 MSSR discriminant analysis

Assuming we have data from $q$ known classes. We let $\boldsymbol{w}_{chj} = \left(\boldsymbol{w}_{hj1}^T, ..., \boldsymbol{w}_{hjm_{hj}}^T\right)^T$ where $\boldsymbol{w}_{chj}$ for $j = 1, ..., n_h$ is one of $n_h$ observed samples from class $h$ $(h = 1, ..., q)$ consisting of $m_{hj}$ points of data. The problem of supervised classification can be described as the problem of correctly assigning a new observation $\boldsymbol{w}_c = \left(\boldsymbol{w}_1^T, ..., \boldsymbol{w}_m^T\right)^T$ to one of the $q$ classes of surfaces, where $\boldsymbol{w}_c$ is a set of $m$ points from a surface belonging to one of the $q$ classes of surfaces. We can achieve a solution to this problem by using MSSR as described in **Section 3**.

Firstly, we suppose that the $h$th class of surfaces can be approximated by a mixture of $g_h$ components whereby, conditioned on $\boldsymbol{w}_{chj}$ belonging to component $i$ $(i = 1, ..., g_h)$, the surface can be approximated using (27). Here $g_h$ can be chosen using a forward selection procedure according to the BIC.

Using the $n_h$ samples from class $h$ $(h = 1, ..., q)$ and the MSSR estimation method from **Section 3**, the probability density function for the $h$th class corresponding to (29) is given as

$$f_h\left(\boldsymbol{y}; \hat{\boldsymbol{\Psi}}_h, \boldsymbol{x}\right) = \sum_{i=1}^{g_h} \pi_i \phi\left(\boldsymbol{y}; \boldsymbol{S}\boldsymbol{\beta}_{hi}, \xi_{hi}^2 \boldsymbol{S}\boldsymbol{S}^T + \sigma_h^2 \boldsymbol{I}_m\right), \tag{35}$$

where $\hat{\boldsymbol{\Psi}}_h = \left(\hat{\sigma}_h^2, \hat{\boldsymbol{\beta}}_{h1}^T, ..., \hat{\boldsymbol{\beta}}_{hg_h}^T, \hat{\xi}_1^2, ..., \hat{\xi}_{g_h}^2, \hat{\boldsymbol{\pi}}_h^T\right)^T$ is the MLE of the parameter vector under the MSSR model of the $h$th class of surfaces, $\hat{\boldsymbol{\pi}}_h = (\hat{\pi}_1, \hat{\pi}_2, ..., \hat{\pi}_{g_h})^T$ and $\boldsymbol{S}$ are the basis evaluations of the observations from $\boldsymbol{w}_c$ as per (6). The MLE can be obtained by applying the EM algorithm as described in the **Appendix**.

If we let the prior probability of observation $\boldsymbol{w}_c$ belonging to class $h$ $(h = 1, ..., q)$ be $\nu_h$, then we can estimate $\nu_h$ by

$$\hat{\nu}_h = \frac{n_h}{\sum_{h=1}^{q} n_h}. \tag{36}$$

Combining the estimates $\hat{\nu}_h$ with the conditional density functions $f_h$, we can estimate the *a posteriori* probability of observation $\boldsymbol{w}_c$ belonging to the class of surfaces $h$ $(h = 1, ..., q)$ by $\hat{\omega}_h$, where

$$\hat{\omega}_h = \frac{\hat{\nu}_h f_h \left(\boldsymbol{y}; \hat{\boldsymbol{\Psi}}_h, \boldsymbol{x}\right)}{\sum_{h'=1}^{q} \hat{\nu}_{h'} f_{h'} \left(\boldsymbol{y}; \hat{\boldsymbol{\Psi}}_{h'}, \boldsymbol{x}\right)}. \tag{37}$$

If we let $\hat{h}$ be the class from which $\boldsymbol{w}_c$ is drawn from, we can use the plug-in Bayes' rule

$$r\left(\boldsymbol{w}_c; \hat{\boldsymbol{\Psi}}_1, ..., \hat{\boldsymbol{\Psi}}_q\right) = \underset{h=1,...,q}{\arg\max}\ \hat{\omega}_h \tag{38}$$

and set $\hat{h} = r\left(\boldsymbol{w}_c; \hat{\boldsymbol{\Psi}}_1, ..., \hat{\boldsymbol{\Psi}}_q\right)$.

In the next section, we provide examples and examine the performance of both MSSR as a surface clustering method and MSSRDA as a classification method.

## 5 Applications and discussions

We now demonstrate the use of SSR, MSSR, and MSSRDA in off-line handwriting recognition problems. The SSR procedure can be shown useful for image compression of handwriting data, MSSR can be used for recovery of images with missing pixels, and MSSRDA can be used for classification of handwritten character images in situations with and without missing data. All of our implementations of SSR, MSSR, and MSSRDA were performed through the **R** software environment (R Development Core Team, 2011).

### *5.1 Handwriting data*

The data set used in our applications is the ZIP code data set from Hastie et al. (2009) and is a subset of a much larger MNIST data set of handwritten digits acquired from the National Institute of Standards and Technology (Le Cun et al., 1990). The data set contains 9298 handwritten Hindu-Arabic numerals, split into a training set and a testing set. The training set contains 7291 observations and consists of 1194 zeros, 1005 ones, 731 twos, 658 threes, 652 fours, 556 fives, 664 sixes, 645 sevens, 542 eights and 644 nines. The testing set contains 2007 observations and consists of 359 zeros, 264 ones, 198 twos, 166 threes, 200 fours, 160 fives, 170 sixes, 147 sevens, 166 eights and 177 nines.

Each observation is a normalized and deslanted 16 by 16 pixels greyscale image where each pixel has an intensity value between -1 and 1. The intensities are normalized values from a discrete range such that -1 is perfect black and 1 is perfect white.

### *5.2 Image compression by SSR*

We can interpret each image from the training set as a set of 7291 image vectors $\boldsymbol{w}_{cj}$, whereby $j = 1, ..., 7291$. For each $j$, the observation is given by $\boldsymbol{w}_{cj} = \left(\boldsymbol{w}_{j1}^T, ..., \boldsymbol{w}_{jm_j}^T\right)^T$, where $m_j = 256$, and the pixel locations are represented by

$$
\begin{array}{llll}
\boldsymbol{x}_{j1} = (1,1) & \boldsymbol{x}_{j2} = (1,2) & \cdots & \boldsymbol{x}_{j16} = (1,16) \\
\boldsymbol{x}_{j17} = (2,1) & \boldsymbol{x}_{j18} = (2,2) & \cdots & \boldsymbol{x}_{j32} = (2,16) \\
\boldsymbol{x}_{j33} = (3,1) & \boldsymbol{x}_{j34} = (3,2) & \cdots & \boldsymbol{x}_{j48} = (3,16) \\
\vdots & \vdots & & \vdots \\
\boldsymbol{x}_{j241} = (16,1) & \boldsymbol{x}_{j242} = (16,2) & \cdots & \boldsymbol{x}_{j256} = (16,16)
\end{array}.
$$

Let $y_{jk}$ be the intensity value at pixel $k$ from observation $j$ $(j = 1, ..., 7291; k = 1, ..., 256)$.

Since, the observations are 16 by 16 pixel images, it can be seen that every point $\boldsymbol{x}_{jk}$ lies in the rectangular domain $R = [1, 16] \times [1, 16]$. In order to fit an intensity surface to each observation, we can fit a $d_1 = 8$ by $d_2 = 8$ spatial spline surface by using a set of $d = 64$ NBFs with centers at $\boldsymbol{c}_1, ..., \boldsymbol{c}_{64}$ as in (1).

Using the SSR procedure from **Section 2**, we fit the estimated spline function $\sum_{l=1}^{d} \hat{\beta}_{jl} s\left(\boldsymbol{x}; \boldsymbol{c}_l\right)$, where $\hat{\boldsymbol{\beta}}_j = \left(\hat{\beta}_{j1}, ..., \hat{\beta}_{jd}\right)^T$ is the vector of estimated basis coefficients for each $j$. The intensities $y_{jk}$ for each observation $\boldsymbol{w}_{cj}$ can be approximated with the vector $\hat{\boldsymbol{\beta}}_j$ by computing the estimate

$$
\hat{y}_{jk} = \sum_{l=1}^{d} \hat{\beta}_{jl} s\left(\boldsymbol{x}_{jk}; \boldsymbol{c}_l\right). \tag{39}
$$

We can consider the estimation process as an image compression since the original observation $\boldsymbol{w}_{cj}$ can be compressed to the set of basis coefficients $\hat{\boldsymbol{\beta}}_j = \left(\hat{\beta}_{j1}, ..., \hat{\beta}_{jd}\right)^T$, which can be used to approximately recover the intensity of each pixel given the correct computational mechanisms. Using $d = 64$, the original 256 pixel image can be compressed by a factor of 4.

The amount of loss due to compression for each image $j$ can be measured by the sample mean squared error (MSE) and the sample mean absolute deviation (MAD) given as

$$\mathrm{MSE}_j = \frac{\sum_{k=1}^{m_j} \left(y_{jk} - \hat{y}_{jk}\right)^2}{m_j} \tag{40}$$

and

$$\mathrm{MAD}_j = \frac{\sum_{k=1}^{m_j} \left|y_{jk} - \hat{y}_{jk}\right|}{m_j}, \tag{41}$$

respectively.

The average MSE and MAD values are reported in (**Table 1**) alongside the naive and spatial information-free predictor

$$\hat{y}_{jk} = \frac{\sum_{k=1}^{m_j} y_{jk}}{m_j} \tag{42}$$

for each observation $\boldsymbol{w}_{cj}$, and the four pixel-averaging process

$$\hat{y}_{jk} = \begin{cases} \frac{y_{jk}+y_{j(k+1)}+y_{j(k+16)}+y_{j(k+17)}}{4} & \text{if } k = 32t + r, \\ \frac{y_{j(k-1)}+y_{jk}+y_{j(k+15)}+y_{j(k+16)}}{4} & \text{if } k = 32t + r + 1, \\ \frac{y_{j(k-16)}+y_{j(k-15)}+y_{jk}+y_{j(k+1)}}{4} & \text{if } k = 32t + r + 16, \\ \frac{y_{j(k-17)}+y_{j(k-16)}+y_{j(k-1)}+y_{jk}}{4} & \text{if } k = 32t + r + 17, \end{cases} \tag{43}$$

where $t = 0, ..., 7$ and $r = 1, 3, 5, ..., 15$ for each observation $\boldsymbol{w}_{cj}$.

From **Table 1** we can compare the compression loss for this data set due the SSR representation of each image with the naive predictor as a baseline. We can also compare the compression loss of SSR against the four pixel-averaging process which also compresses each image by a factor of 4. It can be observed that both the SSR method and the pixel-averaging method are able to use the spatial information to reduce the compression loss when compared to the naive predictor when comparing the MSE and MAD. It can also be observed that SSR is able to reduce the MSE from pixel-average down from a factor of 0.149 to 0.106 whereas it is inferior to averaging by a factor of only 0.008 if comparing the MAD results. This shows that SSR is potentially better at loss reduction than pixel-averaging for the same four fold compression ratio. Along with this result, it can also be noted that although SSR compresses the representation of each image by a factor of four, it is still able to assign a unique intensity to each of the 256 pixels whereas pixel-averaging forces local squares of four pixels to share intensities.

| Measure | MSE | MAD |
|---|---|---|
| Averaging | 0.149 | 0.220 |
| Naive | 0.557 | 0.631 |
| SSR | 0.106 | 0.228 |

Table 1
Average mean squared error and mean absolute deviation for SSR compressed training set images.

The ability to give each individual pixel a unique value allows the SSR compressed images to capture the local curvature of the numerals better than pixel-averaging is able to. We can see this in **Figure 2**. Note that the ability to capture local curvature information appears to come at a cost to background color matching. We can see that the background of the SSR compressed images do not retain the same contrast as the pixel-averaging compressed images when compared to the originals. These are plausible reasons as to why SSR achieves a slightly higher MAD than pixel-averaging but a considerably lower MSE.

### *5.3 Image recovery by MSSR*

We now consider the problem of recovering signals from images with missing pixels. To do this, firstly consider that the training data set can be given as $\boldsymbol{w}_{chj}$, where $\boldsymbol{w}_{chj} = \left(\boldsymbol{w}_{hj1}^T, ..., \boldsymbol{w}_{hjm_j}^T\right)^T$ $(h = 0, ..., 9;\ j = 1, ..., n_h)$ is the $j$th observation of digit $h$ and $n_h$ is the number of images of digit $h$. To demonstrate the missing-data recovery capabilities of MSSR, we consider various levels of missingness defined as $M$, where $M$ is a proportion of pixels with missing observations ($M = 0.5, 0.75, 0.9, 0.95$). For each proportion $M$, we simulated a missing-values data set $\boldsymbol{w}_{chj}^{(M)}$ where $\boldsymbol{w}_{chj}^{(M)} = \left(\boldsymbol{w}_{hj1}^{(M)^T}, ..., \boldsymbol{w}_{hjm_j^{(M)}}^{(M)^T}\right)^T$ is the $j$th observation of digit $h$ such that each $\boldsymbol{w}_{hjk}^{(M)}$ is a uniformly sampled point from $\boldsymbol{w}_{chj}$ $\left(k = 1, ..., m_j^{(M)}\right)$, and $m_j^{(M)} = \left\lfloor (1 - M)\, m_j^{(M)} \right\rfloor$. The floor operator ensures that for each $M$, the missing-data observations have missingness of at least proportion $M$.

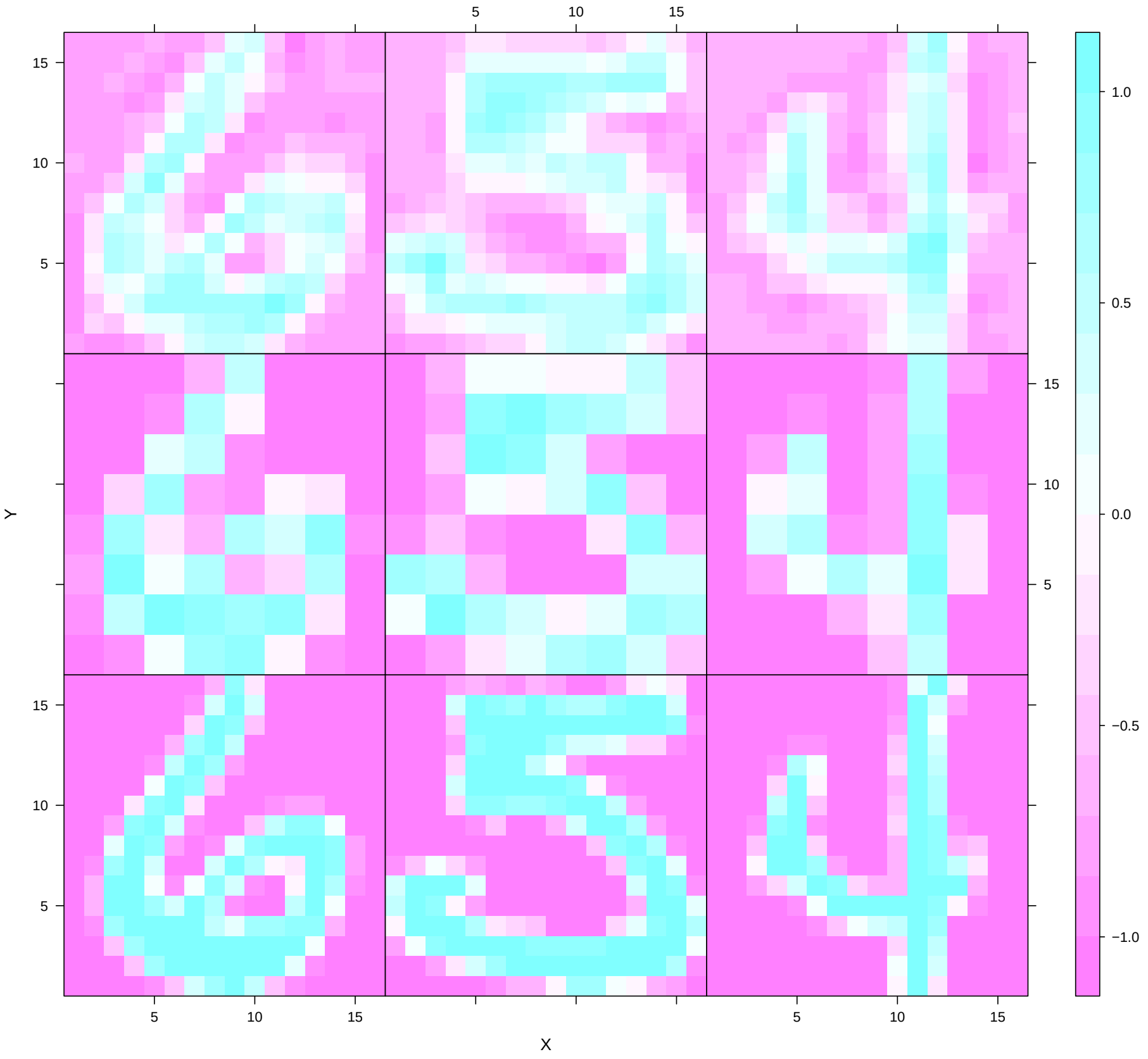

Figure 2. The bottom row shows the true images for the first, second and third numerals from the training data. The middle row shows the same numerals as fitted by pixel-averaging and the top row shows the same images as fitted by SSR.

As in **Section 4**, we assume that conditioned on its membership of the $i$th component of the mixture model, $\boldsymbol{w}_{chj}^{(M)}$ for each $m$ are such that each observed intensity $y_{hjk}$, has the form

$$y_{hjk} = \sum_{l=1}^{d} (\beta_{hi} + b_{hij})\, s(\boldsymbol{x}_{hjk}; \boldsymbol{c}_l) + e_{hjk}, \tag{44}$$

where $i = 1, ..., g_h$, $b_{hij} \sim N(\mathbf{0}_d, \xi_{hi}^2 \boldsymbol{I}_d)$ and $e_{hjk} \sim N(0, \sigma^2)$.

For each digit $h$ and missingness level $M$ ($h = 0, ..., 9; M = 0.5, 0.75, 0.9, 0.95$), we use the MSSR procedure to fit the $g_h$ classes of surfaces where $g_h$ is chosen by BIC. With the resulting model estimates, the original images $\boldsymbol{w}_{chj}$ can be approximately recovered from the images observed with missingness $\boldsymbol{w}_{chj}^{(M)}$ by estimating each pixel $k$ of observation $j$ with

$$\hat{y}_{hjk} = \sum_{l=1}^{d} \left(\hat{\beta}_{h\hat{i}_{hj}l} + \hat{b}_{h\hat{i}_{hj}jl}\right) s(\boldsymbol{x}_{hjk}; \boldsymbol{c}_l), \tag{45}$$

where

$\hat{\boldsymbol{\beta}}_{h\hat{i}_{hj}} = \left(\hat{\beta}_{h\hat{i}_{hj}1}, ..., \hat{\beta}_{h\hat{i}_{hj}d}\right)^T$ and $\hat{\boldsymbol{b}}_{h\hat{i}_{hj}j} = \left(\hat{b}_{h\hat{i}_{hj}j1}, ..., \hat{b}_{h\hat{i}_{hj}jd}\right)^T$ are as described in **Section 3** but now fitted sepearately for each class with the additional leading subscript $h$ as in **Section 4**, and $\hat{i}_{hj}$ is the Bayes-optimal cluster allocation of observation $\boldsymbol{w}_{chj}^{(M)}$. Here $\hat{i}_{hj}$ is the cluster allocation of $\boldsymbol{w}_{chj}^{(M)}$ according to the plug-in Bayes' rule $r\left(\boldsymbol{w}_{chj}^{(M)}; \hat{\boldsymbol{\Psi}}_h\right)$, where

$$r\left(\boldsymbol{w}_{chj}^{(M)}; \hat{\boldsymbol{\Psi}}_h\right) = \arg\max_{i=1,...,g_h} \hat{\tau}_{hij},$$

$\hat{\tau}_{hij}$ is the conditional expectation of numeral $j$ of digit $h$ belonging to class $i$ as described in **Section 3** ($h = 0, ..., 9; i = 1, ..., g_h; j = 1, ..., 7291$), and $\hat{i}_{hj} = r\left(\boldsymbol{w}_{chj}^{(M)}; \hat{\boldsymbol{\Psi}}_h\right)$.

In order to make comparisons, we also consider a number of competing missing-value imputation methods.

### 5.4 Missing-data imputation

We suppose that each set of data with missing elements $\boldsymbol{w}_{ch1}^{(M)}, ..., \boldsymbol{w}_{chn_h}^{(M)}$ can be represented as the matrix

$$A_h^{(M)} = [a_{jk}]_{n_h \times 256}\,, \tag{46}$$

where

$$a_{jk} = \begin{cases} \text{NA} & \text{if } (\boldsymbol{x}_{jk}, y_{jk}) \text{ is missing,} \\ y_{jk} & \text{otherwise,} \end{cases} \tag{47}$$

and NA (not available) is a missing value ($h = 0, ..., 9$; $j = 1, ..., n_h$; $k = 1, ..., 256$).

With each matrix representation $A_h^{(M)}$, we can complete the matrix by computing the estimated $\hat{y}_{jk}$ for each $a_{jk}$ that is equal to NA. The matrix completion techniques that we apply as competition to MSSR are the column mean (CM), $k$-nearest neighbors ($k$-NN), and singular value decomposition (SVD) techniques from Troyanskaya et al. (2001) and the singular-value threshold (SVT) technique (Cai et al., 2010). Using the experimental results from Troyanskaya et al. (2001) and Cai et al. (2010), we chose $k = 1$ and $k = 10$ for the $k$-NN imputation method, and we implemented experimentally optimal thresholds for the SVD and SVT techniques. We implemented these methods through the **imputation** package (Wong, 2011) for **R**.

In **Table 2**, we present the MSE and MAD results for the MSSR and the five competing imputation methods that we have discussed. We note that the results for MSSR are reported as the mean over five runs of the EM algorithm.

The results from **Table 2** show that MSSR generally outperforms the competing methods for image recovery in high-missingness scenarios. We notice that MSSR performs the best on average for the $M = 0.9$ and $M = 0.95$ scenarios with regards to both MSE and MAD, and performs the best in the $M = 0.75$ scenario in terms of the MSE statistic. In the instances where MSSR is not the best method, its results are often highly competitive with those of the best methods. Overall, this shows that MSSR can be a useful tool for image recovery when the images have a high number of missing pixels or low numbers of missing pixels. The MSSR procedure is therefore potentially useful in any situation where missing-data occurs within a set of similar images, such as in the problem of frame repair in motion pictures (Kokaram et al., 1995).

| Missingness ($M$) | 0.5 | | 0.75 | | 0.9 | | 0.95 | |
|---|---|---|---|---|---|---|---|---|
| Measure | MSE | MAD | MSE | MAD | MSE | MAD | MSE | MAD |
| MSSR | 0.095 | 0.15 | **0.168** | 0.244 | **0.247** | **0.323** | **0.298** | **0.363** |
| Column Mean | 0.16 | 0.192 | 0.241 | 0.288 | 0.293 | 0.347 | 0.313 | 0.368 |
| 1-NN | 0.125 | 0.115 | 0.323 | 0.251 | 0.539 | 0.385 | 0.582 | 0.413 |
| 10-NN | 0.077 | **0.108** | 0.172 | **0.216** | 0.293 | 0.332 | 0.324 | 0.365 |
| SVD | **0.076** | 0.124 | 0.202 | 0.259 | 0.29 | 0.344 | 0.317 | 0.37 |
| SVT | 0.147 | 0.184 | 0.234 | 0.285 | 0.29 | 0.347 | 0.312 | 0.369 |

Table 2
Average mean squared error and mean absolute deviation for MSSR, CM, 1-NN, 10-NN, SVD and SVT as applied to the image recovery problem. The best performing method for each measurement and level of missingness is bold.

In **Figures 3** and **4**, we can see the recovered images of the first numeral from the training set through applications for MSSR and the competing methods, respectively. Comparing the recovered images with the original in **Figure 2**, we can see that MSSR is much better able to retain the smoothness between pixels when compared to the images recovered by the other methods. The recovered images by the competing methods often appear to have many spurious pixels which deviate from the original shape of the numeral, whereas the MSSR image is able to recognize the general shape of the numeral and recover an image that has the appropriate curvature and form. The two figures allude to reasons why MSSR achieved lower MSE and MAD in the high missingness scenarios since in the $M = 0.9$ and $M = 0.95$ cases the images recovered by the competing methods appear more jagged and unclear when compared to the respective MSSR recovered images.

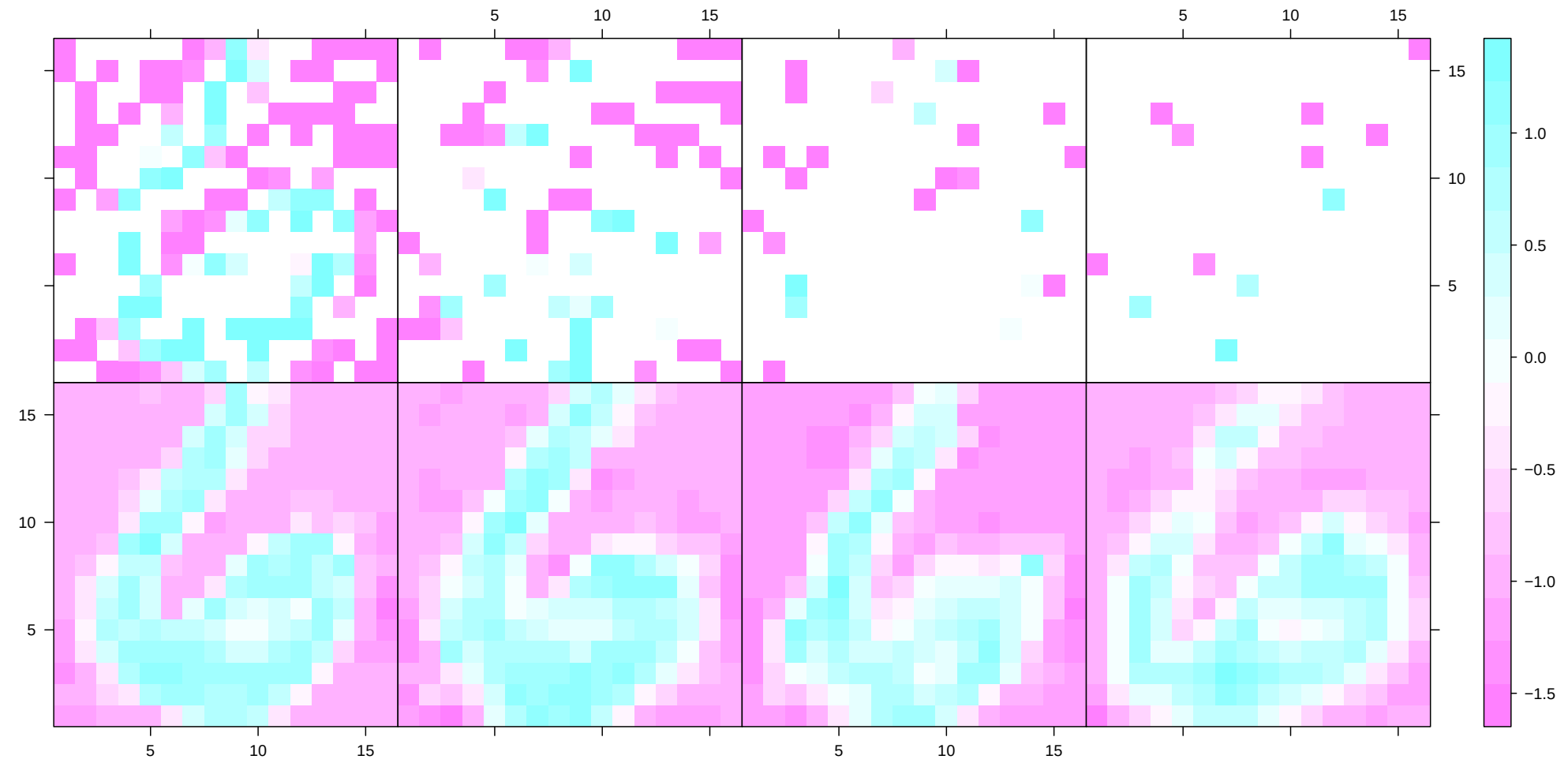

Figure 3. From left to right, the columns correspond to missingness levels of 0.5, 0.75, 0.9 and 0.95 respectively. The first row contains the original missing-data images and the second row contains images that were recovered by MSSR.

In fact, as detailed in **Section 3**, MSSR is also able to recover multiple families of forms of the same digit. Thus, MSSR is not only able to recognize the curvature and form of the number six, but also able to distinguish between multiple varieties of sixes and recover digits based on the most appropriate variety. Examples of the different varieties of sixes found through a single run of MSSR on the $M = 0.5$ training set can be seen in **Figure 5**. These family of sixes were recovered by plotting the functions

$$\sum_{l=1}^{d} \hat{\beta}_{6il} s\left(\boldsymbol{x}; \boldsymbol{c}_l\right) \tag{48}$$

for each $i = 1, ..., g$, where $g = 4$ as selected by BIC forward selection. As noted in **Section**

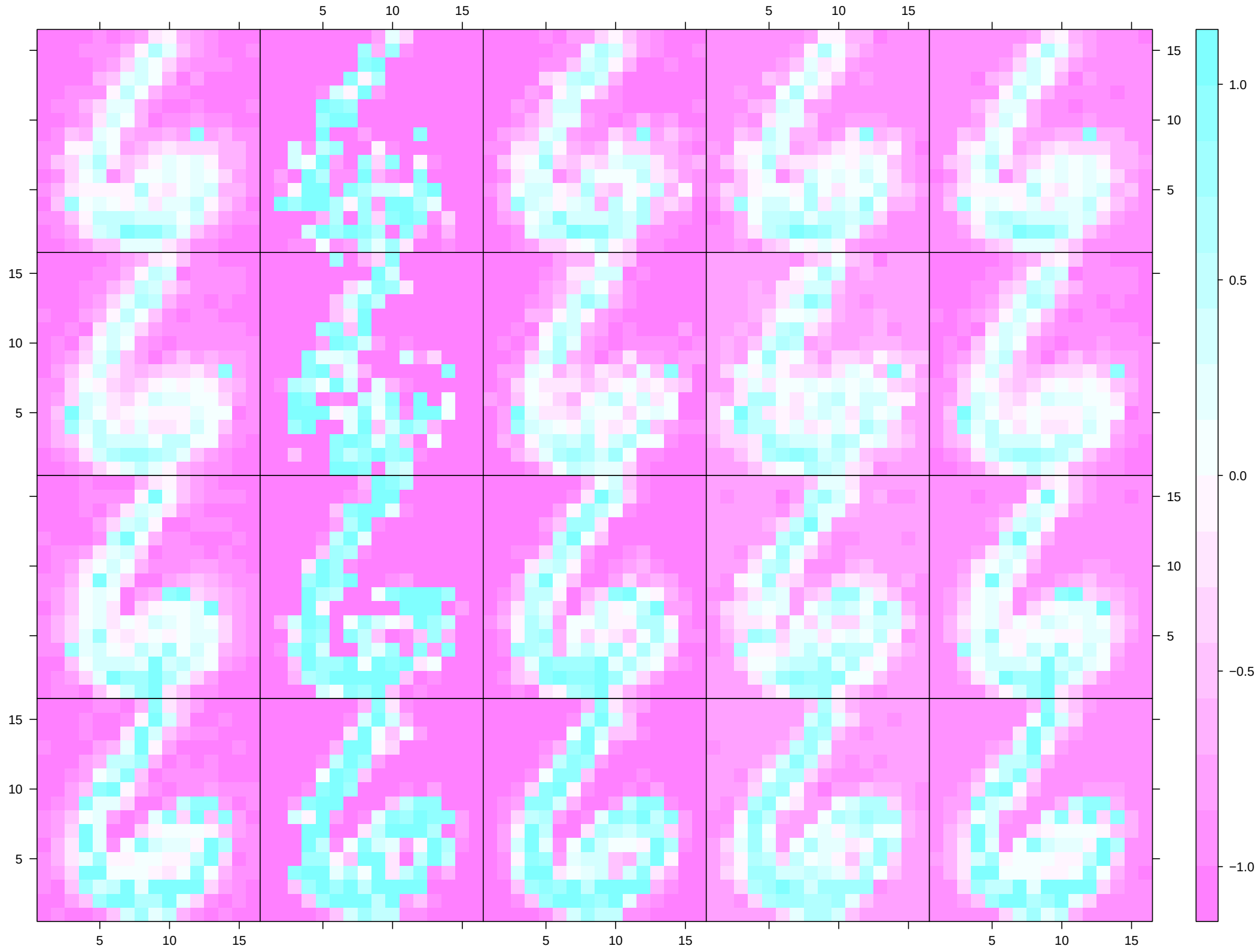

Figure 4. From top to bottom, the rows correspond to missingness levels of 0.95, 0.9, 0.75 and 0.5 respectively. From left to right, the columns contain the images recovered by CM, 1-NN, 10-NN, SVD and SVT.

**3**, these are approximations of the mean functions for which all observed instances of sixes are perturbations. Observing the different families, we can see sixes can be thicker or thinner along the horizontal axis or may have a larger or smaller loop. This recognition would not be possible by assessing the recovered images from the other methods which do not attempt to fit models to images of the same digits.

### *5.5 Classification by MSSRDA*

The MSSR models fitted for image recovery can also be applied to perform handwriting recognition. If we let $\boldsymbol{w}_{c1}^{'(M)}, ..., \boldsymbol{w}_{c2007}^{'(M)}$ be the test set with proportion of missingness $M$, similar to the expressions given in **Section 5.3**, we can perform classification through the MAP rule from **Section 4** as applied to the models fitted as per **Section 5.3**.

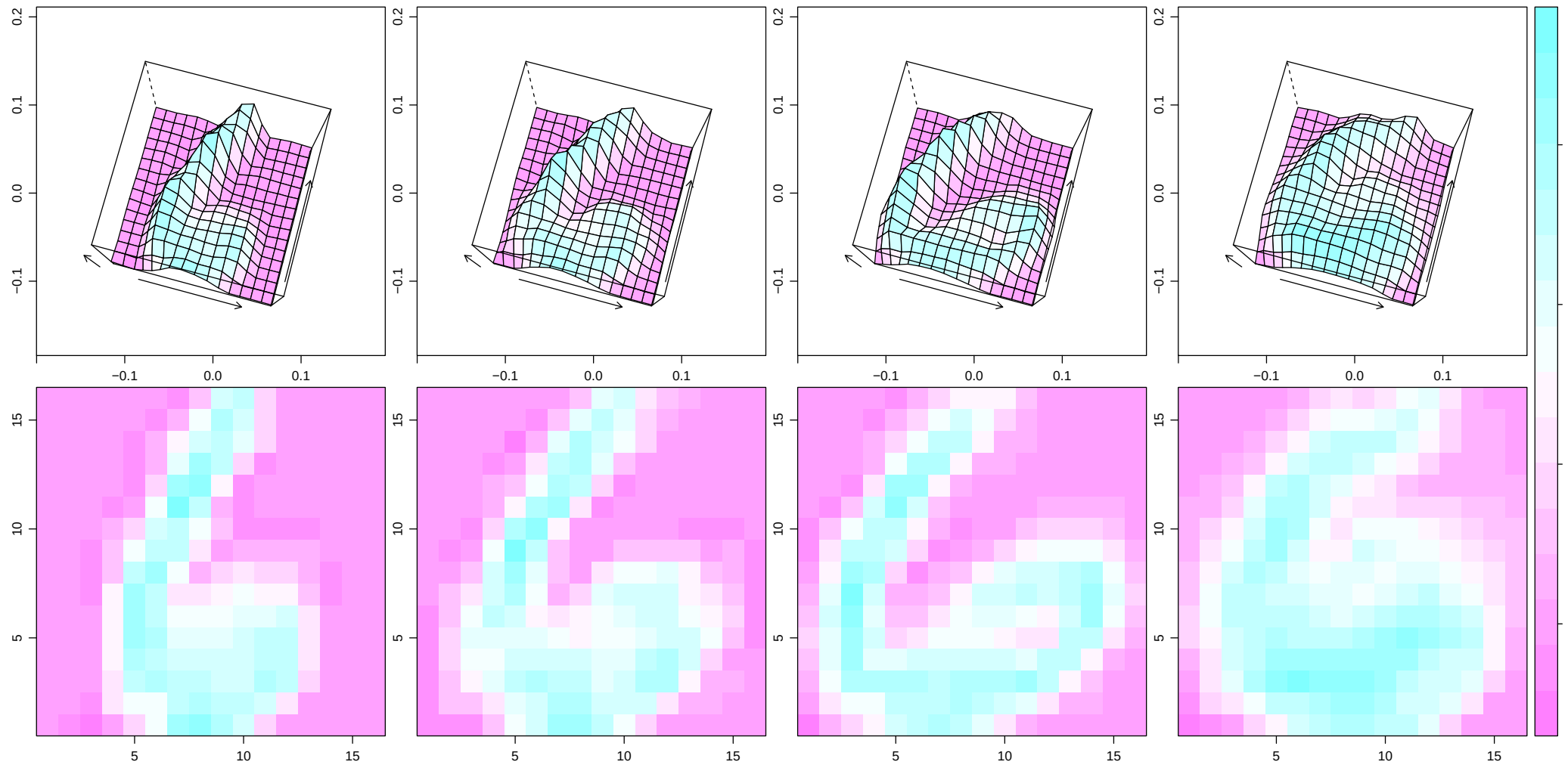

Figure 5. The top row depicts the mean spatial spline surfaces of the four different families of sixes identified within a single run of MSSR using BIC forward selection. The bottom row shows the same four families of sixes as heat maps.

To make comparisons, we also assess the performance of various classification techniques available in the **R**. These techniques include, $k$-nearest neighbors ($k$-NN), linear discriminant analysis (LDA), naive Bayes (NB), and support vector machines (SVM). The $k$-NN classifier was implemented with $k = 1$ and $k = 10$ and the SVM classifier was implemented with both a Gaussian radial basis kernel and a linear kernel. LDA and $k$-NN were implemented through the **MASS** package (Venables and Ripley, 2002), and NB and SVM were implemented through the **e1071** package (Dimitriadou et al., 2000) for **R**. We compared the different methods based on the estimated testing set classification error rate

$$\hat{\text{err}} = \frac{\sum_{j=1}^{n'} l\left(\hat{h}_j, h_j\right)}{n'}, \tag{49}$$

where

$$l\left(\hat{h}_j, h_j\right) = \begin{cases} 1, & \text{if } \hat{h}_j \neq h_j, \\ 0, & \text{otherwise}, \end{cases} \tag{50}$$

$n^{'} = 2007$ is the number of numerals in the testing set, $\hat{h}_j$ is the predicted digit class of observation $\boldsymbol{w}_{cj}^{'(M)}$, $h_j$ is the true digit class for numeral $j$ and $j = 1, ..., n^{'}$.

To apply these methods we need to perform data-imputation to complete the missing-data observations $\boldsymbol{w}_{chj}^{(M)}$ and $\boldsymbol{w}_{cj'}^{'(M)}$ for each $M$ $(h = 0, ..., 9; j = 1, ..., 7291; j' = 1, ..., 2007)$. The training observations $\boldsymbol{w}_{chj}^{(M)}$ can be completed by using the matrix representation and matrix completion techniques discussed in **Section 5.4**. The test set $\boldsymbol{w}_{cj}^{'(M)}, ..., \boldsymbol{w}_{c2007}^{'(M)}$ can be completed by representation as

$$A^{'(M)} = [a_{jk}]_{2007 \times 256}, \tag{51}$$

where

$$a_{jk} = \begin{cases} \text{NA}, & \text{if } (\boldsymbol{x}_{jk}, y_{jk}) \text{ is missing}, \\ y_{jk}, & \text{otherwise}, \end{cases} \tag{52}$$

$j = 1, ..., 2007$ and $k = 1, ..., 256$. Here the matrices $A^{'(M)}$ were not separated into the different digit subgroups. Keeping the numerals together guarantees that the matrix completion algorithms do not obtain extra information about each observation's class.

In **Table 3**, we report the error rate estimate $\hat{\text{err}}$ for MSSRDA and the competing methods over the 25 different scenarios of missingness in the training and testing data sets. Since the competing methods were applied to training and test data that were completed using all five of the methods from **Section 5.4**, it was not feasible to report the classification rates for each individual combination. For brevity, we only report the minimum error rate estimate across all data-imputation methods for each of the competing classifiers. We also note that the results for MSSRDA are reported as the mean over five runs of the EM algorithm.

The MSSRDA procedure appears to be competitive when compared to the competing methods across all the various missingness scenarios. We see that MSSRDA appears to have the greatest advantage over the alternative methods in scenarios where there is high missingness in either the training set, testing set or both. We notice that when the missingness levels are high, such as when $M = 0.9$ or $M = 0.95$ in the training and testing sets, the error rates consistently remained lower than 0.5, which is impressive considering that this is a ten-class discriminant analysis problem. MSSRDA's ability to handle high levels of missing data without imputation is a strong advantage when compared to the other methods since data imputation is often a computationally intensive exercise and choosing the right imputation method for any given classifier is a non-trivial problem in itself. Further, the classification of images with high levels of missingness can be seen as an instance of sparsely sampled functional data. In these sparse data scenarios, MSSRDA is able to borrow information across the large number of sparsely sampled surfaces allowing it to better fit the generative models of each of the classes as previously noted in the one dimensional analogue (James and Sugar, 2003).

From the first row of **Table 3**, we can see the results of MSSRDA and the competing methods on the original handwritten digit recognition problem without artificial missingness. Although the error rates achieved were low, it must be noted that they are significantly higher than the state of the art classification rates for the handwritten digits recognition problem. As collated in Chen et al. (2011), the state of the art methods for classification of the MNIST data set have error rates that are lower than 0.01. However, unlike the MSSRDA and competing methods considered here, the state of the art classifiers such as convolution neural networks (Le Cun et al., 1990) were trained using heavily modified data sets that involved various transforma-tions of the image features and the creation of extra data to supplement the original data set. Because of these elements, these methods are very specific to the task of handwritten digit recognition and therefore are not readily portable to other classification problems. Our method on the other hand required no modifications to the data sets and even when there was missing-data, it required no additional imputation steps. In addition to this, MSSRDA is also highly interpretable via its estimated parameters since it is essentially an application of MSSR to the different digit classes in order to recover the generative families of surfaces from which each image is generated, as demonstrated in **Figure 5**. This is a feature of MSSRDA that is not present in the $k$-NN and SVM classifiers since they do not attempt to recover the generative models. Our method is also more flexible than LDA and NB which are only capable of modeling each digit class as a single generative image.

| Training $M$ | Test $M$ | MSSRDA | LDA | NB | 1-NN | 10-NN | SVM (linear) | SVM (radial) |
|---|---|---|---|---|---|---|---|---|
| 0 | 0 | 0.111 | 0.115 | 0.262 | **0.056** | 0.063 | 0.070 | 0.062 |
| 0 | 0.5 | 0.120 | 0.142 | 0.272 | **0.066** | 0.084 | 0.105 | 0.070 |
| 0 | 0.75 | **0.146** | 0.255 | 0.449 | 0.149 | 0.148 | 0.259 | 0.172 |
| 0 | 0.9 | **0.261** | 0.599 | 0.809 | 0.486 | 0.445 | 0.672 | 0.622 |
| 0 | 0.95 | **0.408** | 0.791 | 0.851 | 0.699 | 0.669 | 0.838 | 0.834 |
| 0.5 | 0 | 0.135 | 0.116 | 0.203 | **0.057** | 0.060 | 0.078 | 0.073 |
| 0.5 | 0.5 | 0.142 | 0.148 | 0.24 | **0.071** | 0.089 | 0.103 | 0.082 |
| 0.5 | 0.75 | 0.175 | 0.264 | 0.453 | **0.148** | **0.148** | 0.226 | 0.166 |
| 0.5 | 0.9 | **0.284** | 0.649 | 0.805 | 0.481 | 0.444 | 0.646 | 0.58 |
| 0.5 | 0.95 | **0.414** | 0.802 | 0.855 | 0.756 | 0.710 | 0.800 | 0.788 |
| 0.75 | 0 | 0.132 | 0.112 | 0.186 | **0.076** | 0.090 | 0.089 | 0.111 |
| 0.75 | 0.5 | 0.142 | 0.146 | 0.232 | 0.116 | 0.118 | **0.112** | 0.118 |
| 0.75 | 0.75 | **0.168** | 0.246 | 0.468 | 0.174 | 0.175 | 0.205 | 0.187 |
| 0.75 | 0.9 | **0.276** | 0.623 | 0.75 | 0.53 | 0.502 | 0.593 | 0.558 |
| 0.75 | 0.95 | **0.414** | 0.805 | 0.849 | 0.796 | 0.784 | 0.762 | 0.769 |
| 0.9 | 0 | 0.141 | 0.136 | 0.177 | **0.139** | 0.147 | 0.140 | 0.179 |
| 0.9 | 0.5 | **0.148** | 0.167 | 0.225 | 0.170 | 0.170 | 0.163 | 0.197 |
| 0.9 | 0.75 | **0.183** | 0.251 | 0.495 | 0.217 | 0.216 | 0.224 | 0.267 |
| 0.9 | 0.9 | **0.288** | 0.557 | 0.793 | 0.535 | 0.523 | 0.541 | 0.534 |
| 0.9 | 0.95 | **0.426** | 0.775 | 0.857 | 0.777 | 0.764 | 0.742 | 0.738 |
| 0.95 | 0 | 0.159 | 0.163 | 0.192 | 0.168 | 0.166 | **0.153** | 0.197 |
| 0.95 | 0.5 | **0.162** | 0.170 | 0.223 | 0.182 | 0.183 | 0.182 | 0.218 |
| 0.95 | 0.75 | **0.19** | 0.243 | 0.492 | 0.231 | 0.228 | 0.251 | 0.333 |
| 0.95 | 0.9 | **0.31** | 0.575 | 0.828 | 0.556 | 0.551 | 0.546 | 0.606 |
| 0.95 | 0.95 | **0.435** | 0.779 | 0.876 | 0.781 | 0.775 | 0.729 | 0.749 |

Table 3
Average test set classification error rates for the MSSRDA as well as error rates for LDA, NB, 1-NN, 10-NN, linear SVM and Gaussian radial SVM. The bold font denotes the method that achieved the lowest error rate in each scenario. Each row presents results for different proportions of missing pixels in the training and test data sets. The proportion of missingness in the training and test data are displayed in the first and second columns, respectively.

## 6 Conclusions

The literature on the problem of clustering and classifying functional data has mainly been focused on the analysis of univariate functions. In this paper, we presented the MSSR and MSSRDA methods to expand the scope of the literature to bivariate functional data. Both the methods presented can be viewed as bivariate extensions to the spline-based univariate functional data clustering and classification methods developed in James and Hastie (2001) and James and Sugar (2003), using the SSR model from Malfait and Ramsay (2003), Ramsay et al. (2011), and Sangalli et al. (2013).

We show that the SSR, MSSR, and MSSRDA can all be fitted using maximum likelihood estimation and in the case of MSSR we also provide the appropriate EM algorithm steps in order to compute the MLE. Discrimination by MSSRDA was shown to simply involve multiple applications of MSSR to different classes and can be viewed as the mixture discriminant analysis (Hastie and Tibshirani, 1996) analog of the functional LDA method in James and Hastie (2001).

In order to show the potential of our methodologies, we have applied SSR, MSSR, and MSSRDA to problems of image compression, image recovery, and handwritten character recognition, respectively. The application was conducted using the ZIP code data set from Hastie et al. (2009).

We found that in the problem of image compression, SSR was able to achieve highly competitive loss levels when compared to pixel-averaging for a fixed ratio of compression. Moreover, it was observed that SSR was better able to retain the curvature of the numerals in each image.

Artificially generated missing-data was generated to assess the image recovery capabilities of MSSR. It was found that MSSR was not only the most effective method tested for recovering missing-data at high levels of missingness, it was also able to infer multiple varieties of each digit class. This capability of recovering the generative models of a class of images was unique to MSSR amongst the methods explored.

Lastly, we showed that MSSRDA can achieve error rates that are competitive with other classifiers which do not modify the feature space especially when comparing results for situations where the data has a high level of missingness. As well as this, similarly to MSSR, MSSRDA could also recover the generative models of each of the classes of digits which can be used to make inferences instead of simply providing a discriminant function.

There are numerous extensions of this work that could be explored in future. These include the use of higher order NBFs to achieve smoother model fits, defining the methods over irregular, non-rectangular grids, and extending the method to three-dimensional functional data for applications to situations such as spatiotemporal classification problems in fMRI and climatology.

**Appendix**

In this appendix, we detail the EM algorithm for maximizing the likelihood function (30) over the parameter vector $\boldsymbol{\Psi} = \left(\sigma^2, \boldsymbol{\beta}_1^T, ..., \boldsymbol{\beta}_g^T, \xi_1^2, ...\xi_g^2, \boldsymbol{\pi}^T\right)^T$ through the use of the complete-data likelihood function (32). Starting at the value $\boldsymbol{\Psi}^{(0)} = \left(\sigma^{(0)^2}, \boldsymbol{\beta}_1^{(0)T}, ..., \boldsymbol{\beta}_g^{(0)T}, \xi_1^{(0)^2}, ...\xi_g^{(0)^2}, \boldsymbol{\pi}^{(0)T}\right)^T$ for $\boldsymbol{\Psi}$, the E-step of the algorithm on the $(k+1)$ th iteration proceeds by computing the conditional expectations

$$E_{\boldsymbol{\Psi}^{(k)}}\left(Z_{ij}|\boldsymbol{w}_{cj}\right) = \frac{\pi_i^{(k)}\phi\left(\boldsymbol{y}_j; \boldsymbol{S}_j\boldsymbol{\beta}_i^{(k)}, \xi_i^{(k)^2}\boldsymbol{S}_j\boldsymbol{S}_j^T + \sigma^{(k)^2}\boldsymbol{I}_{m_j}\right)}{\sum_{i'=1}^g \pi_{i'}^{(k)}\phi\left(\boldsymbol{y}_j; \boldsymbol{S}_j\boldsymbol{\beta}_{i'}^{(k)}, \xi_{i'}^{2(k)}\boldsymbol{S}_j\boldsymbol{S}_j^T + \sigma^{(k)^2}\boldsymbol{I}_{m_j}\right)}, \tag{53}$$

$$E_{\boldsymbol{\Psi}^{(k)}}\left(\boldsymbol{b}_{ij}|Z_{ij}=1, \boldsymbol{w}_{cj}\right) = \xi_i^{(k)^2}\boldsymbol{S}_j^T\left(\xi_i^{(k)^2}\boldsymbol{S}_j\boldsymbol{S}_j^T + \sigma^{(k)^2}\boldsymbol{I}_{m_j}\right)^{-1}\left(\boldsymbol{y}_j - \boldsymbol{S}_j\boldsymbol{\beta}_i^{(k)}\right), \tag{54}$$

$$E_{\boldsymbol{\Psi}^{(k)}}\left(\boldsymbol{b}_{ij}^T\boldsymbol{b}_{ij}|Z_{ij}=1, \boldsymbol{w}_{cj}\right) = \lambda_{\boldsymbol{b}_{ij}} + \boldsymbol{b}_{ij}^{(k+1)T}\boldsymbol{b}_{ij}^{(k+1)}, \tag{55}$$

$$E_{\boldsymbol{\Psi}^{(k)}}\left(\boldsymbol{e}_{ij}|Z_{ij}=1, \boldsymbol{w}_{cj}\right) = \boldsymbol{y}_j - \boldsymbol{S}_j\left(\boldsymbol{\beta}+\boldsymbol{b}_{ij}^{(k+1)}\right), \tag{56}$$

and

$$E_{\boldsymbol{\Psi}^{(k)}}\left(\boldsymbol{e}_{ij}^T\boldsymbol{e}_{ij}|Z_{ij}=1, \boldsymbol{w}_{cj}\right) = \lambda_{\boldsymbol{e}_{ij}} + \left(\boldsymbol{y}_j - \boldsymbol{S}_j\left(\boldsymbol{\beta}+\boldsymbol{b}_{ij}^{(k+1)}\right)\right)^T\left(\boldsymbol{y}_j - \boldsymbol{S}_j\left(\boldsymbol{\beta}+\boldsymbol{b}_{ij}^{(k+1)}\right)\right), \tag{57}$$

where $\boldsymbol{b}_{ij}^{(k+1)}$, $\lambda_{\boldsymbol{b}_{ij}}$ and $\lambda_{\boldsymbol{e}_{ij}}$ are given as

$$\boldsymbol{b}_{ij}^{(k+1)} = E_{\boldsymbol{\Psi}^{(k)}}\left(\boldsymbol{b}_j | Z_{ij} = 1, \boldsymbol{w}_{cj}\right), \tag{58}$$

$$\lambda_{\boldsymbol{b}_{ij}} = \operatorname{tr}\left(\xi_i^{(k)^2}\left(\boldsymbol{I}_d - \xi_i^{(k)^2}\boldsymbol{S}_j^T\left(\xi_i^{(k)^2}\boldsymbol{S}_j\boldsymbol{S}_j^T + \sigma^{(k)2}\boldsymbol{I}_{m_j}\right)^{-1}\boldsymbol{S}_j\right)\right), \tag{59}$$

and

$$\lambda_{\boldsymbol{e}_{ij}} = \operatorname{tr}\left(\xi_i^{(k)^2}\boldsymbol{S}_j\left(\boldsymbol{I}_d - \xi_i^{(k)^2}\boldsymbol{S}_j^T\left(\xi_i^{(k)^2}\boldsymbol{S}_j\boldsymbol{S}_j^T + \sigma^{(k)2}\boldsymbol{I}_{m_j}\right)^{-1}\boldsymbol{S}_j\right)\boldsymbol{S}_j^T\right), \tag{60}$$

respectively, for each $i$ and $j$.

Letting $\tau_{ij}^{(k)} = E_{\hat{\boldsymbol{\Psi}}^{(k)}}\left(Z_{ij}|\boldsymbol{w}_{cj}\right)$, the M-step on the $(k+1)$ th iteration produces the updated estimates

$$\pi_{ij}^{(k+1)} = \sum_{j=1}^{n}\frac{\tau_{ij}^{(k)}}{n}, \tag{61}$$

$$\boldsymbol{\beta}_i^{(k+1)} = \left(\sum_{j=1}^{n}\tau_{ij}^{(k)}\boldsymbol{S}_j^T\boldsymbol{S}_j\right)^{-1}\left(\sum_{j=1}^{n}\tau_{ij}^{(k)}\boldsymbol{S}_j^T\left(\boldsymbol{y}_j - \boldsymbol{S}_j\boldsymbol{b}_{ij}^{(k+1)}\right)\right), \tag{62}$$

and

$$\xi_i^{(k+1)^2} = \frac{\sum_{j=1}^{n}\tau_{ij}^{(k)}\left(\boldsymbol{b}_{ij}^{(k+1)T}\boldsymbol{b}_{ij}^{(k+1)} + \lambda_{\boldsymbol{b}_{ij}}\right)}{d\sum_{j=1}^{n}\tau_{ij}^{(k)}} \tag{63}$$

for each $i$, and

$$\sigma^{(k+1)^2} = \frac{\sum_{j=1}^{n}\sum_{i=1}^{g}\tau_{ij}^{(k)}\left(\left(\boldsymbol{y}_j - \boldsymbol{S}_j\left(\boldsymbol{\beta}+\boldsymbol{b}_{ij}^{(k+1)}\right)\right)^T\left(\boldsymbol{y}_j - \boldsymbol{S}_j\left(\boldsymbol{\beta}+\boldsymbol{b}_{ij}^{(k+1)}\right)\right) + \lambda_{\boldsymbol{e}_{ij}}\right)}{\sum_{j=1}^{n}m_j}. \tag{64}$$